# High-pressure synthesis and the enhancement of the superconducting properties of FeSe$_{0.5}$Te$_{0.5}$

Mohammad Azam[1], Manasa Manasa[1], Tatiana Zajarniuk[2], Ryszard Diduszko[3], Tomasz Cetner[1], Andrzej Morawski[1], Jarosław Więckowski[2], Andrzej Wiśniewski[2], Shiv J. Singh[1,*]

[1] Institute of High Pressure Physics (IHPP), Polish Academy of Sciences, Sokołowska 29/37, 01-142 Warsaw, Poland
[2] Institute of Physics, Polish Academy of Sciences, Aleja Lotników 32/46, 02-668 Warsaw, Poland
[3] Łukasiewicz Research Network Institute of Microelectronics and Photonics, Aleja Lotników 32/46, 02-668 Warsaw, Poland

* Correspondence: sjs@unipress.waw.pl

**Abstract:** A series of FeSe$_{0.5}$Te$_{0.5}$ bulk samples have been prepared through the high gas pressure and high-temperature synthesis (HP-HTS) method to optimize the growth conditions, for the first time and investigated for their superconducting properties using structural, microstructure, transport, and magnetic measurements to reach the final conclusions. *Ex-situ* and *in-situ* processes are used to prepare bulk samples under a range of growth pressures using Ta-tube and without Ta-tube. The parent compound synthesized by convenient synthesis method at ambient pressure (CSP) exhibits a superconducting transition temperature of 14.8 K. Our data demonstrate that the prepared FeSe$_{0.5}$Te$_{0.5}$ sealed in a Ta-tube is of better quality than the samples without a Ta-tube, and the optimum growth conditions (500 MPa, 600°C for 1 h) are favourable for the development of the tetragonal FeSe$_{0.5}$Te$_{0.5}$ phase. The optimum bulk FeSe$_{0.5}$Te$_{0.5}$ depicts a higher transition temperature of 17.3 K and a high critical current density of the order of >10$^4$ A/cm$^2$ at 0 T, which is improved over the entire magnetic field range and almost twice higher than the parent compound prepared through CSP. Our studies confirm that the high-pressure synthesis method is a highly efficient way to improve the superconducting transition, grain connectivity, sample density, and also pinning properties of a superconductor.

## 1. Introduction

Iron-based superconductor (FBS) was discovered in 2008 [1], and since then numerous compounds have been identified belonging to this family [2, 3]. These high $T_c$ materials can be categorized into six families based on the structure of the parent compounds, such as *RE*FeAsO (*RE*1111; *RE* = rare earth), *A*Fe$_2$As$_2$ (122; *A* = Ba, K, Ca), (Li/Na)FeAs (111), thick perovskite-type oxide blocking layers such as Sr$_4$V$_2$O$_6$Fe$_2$As$_2$ (22456), Sr$_4$Sc$_2$O$_6$Fe$_2$P$_2$ (42622), etc., and chalcogenide Fe*X* representing 11 (*X* = chalcogenide) [4, 5, 6]. 11 is the simplest family of FBS and has the highest $T_c$ of around 15 K for FeSe$_{1-x}$Te$_x$ (*x* = 0.5) samples at ambient pressure [7, 3, 8]. However, FeSe has a complicated phase diagram [9, 10] due to many stable phases such as hexagonal Fe$_7$Se$_8$, monoclinic Fe$_3$Se$_4$, orthorhombic FeSe$_2$, hexagonal δ-Fe$_x$Se and tetragonal β-Fe$_x$Se phase [11]. Interestingly, only the tetragonal structure β-Fe$_x$Se exhibits the superconducting transition of 8 K at ambient pressure [12, 13]. Due to these stable phases, it's always challenging to prepare a completely pure superconducting phase either in single crystal [14, 15] or polycrystalline samples [3, 16, 17, 7]. Some of these stable phases, particularly hexagonal δ-Fe$_x$Se, and hexagonal $Fe_7Se_8$, typically coexist with the predominant tetragonal $β-Fe_xSe$ phase [18] during the growth process and are not suitable for superconducting properties [19, 20, 21, 22, 23]. Recent research has demonstrated that convenient synthesis methods at ambient pressure (CSP) are inefficient for improving the critical current properties, grain connectivity, and phase purity of the bulk samples, as well as for the practical applications [20, 21, 22, 19]. The preparation of Fe(Se,Te) has been reported by melting routes [20, 24], where the pristine Fe(Se,Te) samples were synthesized at a very high heating temperature of 880-1000°C for a long heating time, and various annealing treatments of these Fe(Se,Te) samples were also carried out to influence the sample properties, but it was still impossible to reduce the foreign phases and to obtain high-quality bulks. Therefore, we need to adopt a novel growth approach or procedure in order to improve the sample qualities, reduce the impurity phases during the tetragonal 11-phase formation, and also enhance the superconducting properties of these materials [25].

In the last 15 years, numerous investigations for iron-based superconductors have been reported with the applied external pressure studies where the used samples were prepared through CSP [26, 27]. Under 4.5 GPa of an applied external pressure, the superconducting transition of the 11 family is enhanced up to 37.6 K [13]. Through CSP, there are always issues related to the size and purity of FBS samples [2]. High-pressure synthesis is a useful and effective way to grow the new solid-state phase [28]. A few investigations based on the pressure synthesis of iron-based superconductors confirm that this method can be an effective method to improve the sample's qualities with high superconductivity properties [29, 30, 2]. However, we need more devoted work in this direction [2]. Two kinds of pressure techniques [31] are used for the synthesis/growth process of FBS: *1)* Most studies have employed solid pressure medium [32, 33], such as diamond anvil cells and hot-isostatic pressing, which typically have a sample volume of the order of ~1 cm$^3$, where the pressure medium and heating elements frequently come into contact with the samples during the growth process. These factors make it difficult to prepare long-size crystals or large amounts of powder with a completely pure phase formation [29]. *2)* On the other hand, there is another pressure technique, *i.e.* the hot-isostatic gas pressure technique, which works based on the inert gas pressure medium and has a ~15 cm$^3$ sample space [31, 34]. This technique has been used in a few investigations for cuprate superconductors [34, 31] and one paper concerning FeSe [35]. According to our knowledge, no reports are available based on this synthesis approach for other iron-based superconductors. According to a recent study, FeSe and Cu-doped FeSe exhibit a high transition temperature of up to 37 K at ambient pressure through the pressure quenching at low temperatures [30]. It's interesting to note that this high $T_c$ phase of FeSe remains stable for 7 days in the temperature range from 4 K to 300 K [30] under ambient pressure. These results imply that the growth of FBS through high-pressure synthesis can be an effective approach to retaining the high $T_c$ phase and the enhanced superconducting properties at ambient pressure.

To understand the high-pressure synthesis effects on the superconducting properties of 11 family, we have used the high-pressure and high-temperature gas isostatic method (HP-HTS) [31, 34, 35] to grow the polycrystalline $FeSe_{0.5}Te_{0.5}$ and studied the superconducting properties of these materials in comparison to bulk $FeSe_{0.5}Te_{0.5}$ prepared through the convenient synthesis method at ambient pressure (CSP) [22, 21]. Since no study based on this high-pressure growth is available, various $FeSe_{0.5}Te_{0.5}$ bulks are prepared by *ex-situ* and *in-situ* processes with the sealed Tantalum (Ta) tube or without the Ta-tube under the inert gas pressure effect to optimize the growth process of $FeSe_{0.5}Te_{0.5}$. The hexagonal phase is portrayed by the structural and microstructural analysis as an impurity phase that is extremely sensitive to the growth pressure and synthesis conditions. Interestingly, $FeSe_{0.5}Te_{0.5}$ sample sealed in a Ta-tube prepared at 500 MPa and placed into an HP-HTS chamber for a short heating time of 1 h displays a high $T_c$ of up to 17.2 K and also enhances the critical current density ($J_c$). A significant dependence on high-pressure growth and superconducting properties is discussed concerning the grain connectivity, superconducting transition $T_c$, and the critical current density.

## 2. Experimental details

The chosen composition FeSe$_{0.5}$Te$_{0.5}$ samples were synthesized in two ways using the HP-HTS method, as mentioned in Figure 1. ***Ex-situ:*** In the first step, polycrystalline samples with a nominal composition of FeSe$_{0.5}$Te$_{0.5}$ were prepared by using the convenient solid-state reaction method at ambient pressure (CSP) [22, 21] as mentioned in Figure 1 as the parent compound. The initial precursors, consisting of Fe powder (99.99% purity, Alfa Aesar), Se (99.99% purity, Alfa Aesar), and Te (99.99% purity, Alfa Aesar), were mixed in an agate mortar for 15-20 min by their nominal composition of FeSe$_{0.5}$Te$_{0.5}$. These thoroughly mixed powders were cold-pressed into discs of 10 mm (diameter) under 6 tons of uniaxial pressure, and then they were sealed into an evacuated quartz tube. More details about the synthesis process can be found elsewhere [22, 21]. The evacuated quartz tube is used to reduce the oxygen and moisture atmosphere around the precursors and also helps during the sealing process of the quartz tube. These quartz tubes were heated to 600°C for 11 h in the box furnace. Generally, we prepared 6-7 grams of samples in one batch, and the resulting powders were then mixed once again in an agate mortar. The disk-shaped pellets with a diameter of 10 mm were again prepared. For the second step heating procedure, we have employed some of these pellets directly in the HP-HTS chamber in an open crucible, while other pellets sealed in a Ta-tube are used in the HP-HTS chamber. ARC melting is used to seal the pellets in a Ta-tube in an inert gas atmosphere. ***In-situ:*** In this case, the starting reagents, *i.e.*, Fe powder (99.99% purity, Alfa Aesar), Se (99.99% purity, Alfa Aesar), and Te (99.99% purity, Alfa Aesar), were mixed according to the selected composition of FeSe$_{0.5}$Te$_{0.5}$, which was thoroughly mixed in an agate mortar and then cold-pressed into discs having a diameter of 10 mm. Some pellets are employed directly in an open crucible in the HP-HTS chamber, while others are sealed into a Ta tube through an ARC melter and then placed inside the HP-HTS chamber, as depicted in the synthesis block diagram (Figure 1).

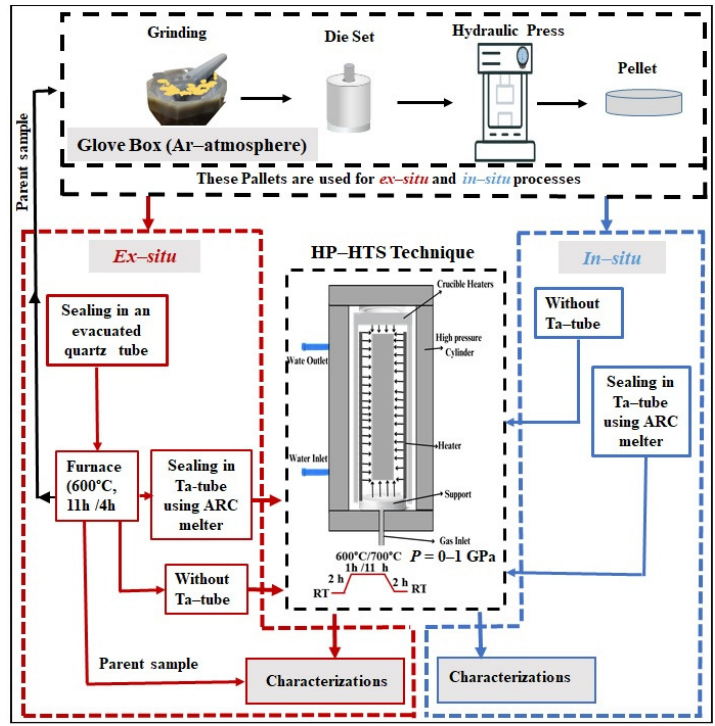

**Figure 1.** Block diagram of the growth process of FeSe$_{0.5}$Te$_{0.5}$ bulks using the HP-HTS method.

We have used the Hot-Isostatic Gas Pressure (HIP) technique, which can generate the inert gas pressures of up to ~1.8 GPa in a cylinder chamber fitted with a furnace capable of reaching 1700°C *i.e.*, an HP-HTS method. This pressure technique allows several cm$^3$ of free volume for the sample's growth, and the stability of temperature can be controlled with an accuracy of 0.5 K through a programmable temperature controller. In all our experiments, we have used the argon gas atmosphere to perform high isostatic gas pressures of up to 1.0 GPa for various reaction times [34]. The internal single- or three-zone Kanthal furnace in our high-pressure technical chamber is part of our experimental equipment, which also includes a 1.8 GPa gas compressor [31, 34, 35]. After HP-HTS growth, the final pellets were compact but not in a disk shape. We have cut these samples into a bar shape for the resistivity and magnetic measurements. To avoid any kind of degradation during the growth process, all chemical manipulations were performed in an argon-filled glove box. This process was used to prepare several samples from different batches, all of which demonstrated high consistency in terms of superconducting properties.

Structural characterizations of all prepared samples were performed by the powder X-ray diffraction method (XRD), which is carried out on the Rigaku SmartLab 3kW diffractometer with filtered Cu-K$\alpha$ radiation (wavelength: 1.5418 Å, power: 30 mA, 40 kV) and a Dtex250 linear detector with the measured profile from 5° to 70° with a very small step of 0.01°/min. We have used Rigaku's PDXL software and the ICDD PDF4+ 2021 standard diffraction patterns database to analyze the lattice parameters, the profile analysis, and the quantitative values of impurity phases (%). Field-emission scanning electron microscope equipped with energy-dispersive X-ray spectroscopy (EDAX) are used for the microstructural characterizations and compositional and elemental analyses of the prepared samples. A vibrating sample magnetometer (VSM) attached to a Quantum Design PPMS is employed for the magnetic measurements up to 9 T in the temperature range of 5-22 K under zero-field and field-cooling conditions. During the zero-field cooled (ZFC) process, the bulk sample was cooled down to 5 K in a zero magnetic field, and then the magnetic field was applied to collect the data during the warming process. A closed-cycle refrigerator was used for the zero-field resistivity measurements in the temperature range from 7 to 300 K at different applied currents (5-20 mA). The sample was cut into a rectangular shape for the resistivity and magnetic measurements, where thin copper wires were used to make the electrical currents with the silver epoxy as the four-probe method for the resistivity measurements, and all data were collected during the warming process with a slow heating rate.

## 3. Results and discussion

*3.1. Structural analysis*

Table 1 shows the list of sample codes and the synthesis conditions of all prepared samples using HP-HTS. The parent sample refers to the composition FeSe$_{0.5}$Te$_{0.5}$, prepared at ambient pressure (CSP), and more details about its basic superconducting properties are reported elsewhere [22, 21]. FeSe$_{0.5}$Te$_{0.5}$ samples prepared by high-pressure techniques sealed into Ta-tubes or without Ta-tubes at different pressures (0-1 GPa) are characterized by X-ray powder diffraction as depicted in Figure 2(a). To be more clear, we have shown the zoom XRD patterns around the main peak (101) of the tetragonal superconducting phase of FeSe$_{0.5}$Te$_{0.5}$ in Figure 2(b). The parent compound FeSe$_{0.5}$Te$_{0.5}$ displays the main tetragonal phase with space group *P4/nmm* as well as a small amount of hexagonal phase (~5-6%) which is consistent with previously reported papers [23, 20, 3]. HIP-S1 to HIP-S5 have been prepared under an applied pressure of 1 GPa. HIP-S1 and HIP-S2 have been used as *ex-situ* synthesis processes, as mentioned in the experimental part and Figure 1, where the samples were sealed into Ta-tubes. HIP-S1 heated at 1 GPa for 1 h depicts the main phase of FeSe$_{0.5}$Te$_{0.5}$ but also observes the enhanced hexagonal phase (~27%) and a small amount of FeSe$_2$ phase. For the enhanced sintering time at 1 GPa up to 11 h, *i.e.,* HIP-S2, has the main hexagonal phase Fe$_7$Se$_8$, and the tetragonal phase of FeSe$_{0.5}$Te$_{0.5}$ was observed as a second phase. It suggests that the long sintering time at 1 GPa of the sealed Fe(Se,Te) sample into the Ta-tube converted the tetragonal phase to the hexagonal phase.

**Table 1.** A list of the samples with synthesis conditions and sample codes. More details about the growth process are mentioned in the experimental part and Figure 1.

| Sample synthesis conditions | Growth Process | Sample's Code |
|---|---|---|
| First step: 600 °C, 11 h, ambient pressure<br>Second step: 600 °C, 4 h at ambient pressure (without Ta-tube) | Parent | Parent |
| First step: 600 °C, 11 h, ambient pressure<br>Second step: 600 °C, 1 h, 1 GPa (with Ta-tube) | *ex-situ* | HIP-S1 |
| First step: 600 °C, 11 h, ambient pressure<br>Second step: 600 °C, 11 h, 1 GPa (with Ta-tube) | *ex-situ* | HIP-S2 |
| First step: 600 °C, 11 h, 1 GPa (with Ta-tube) | *in-situ* | HIP-S3 |
| First step: 700 °C, 1 h, 1 GPa (without Ta-tube) | *in-situ* | HIP-S4 |
| First step: 600 °C, 11 h, 1 GPa (without Ta-tube) | *in-situ* | HIP-S5 |
| First step: 600 °C, 11 h, ambient pressure<br>Second step: 600 °C, 11 h, 700 MPa (without Ta-tube) | *ex-situ* | HIP-S6 |
| First step: 600 °C, 11 h, ambient pressure<br>Second step: 600 °C, 11 h, 1 GPa-750MPa (without Ta-tube) | *ex-situ* | HIP-S7 |
| First step: 600 °C, 11 h, ambient pressure<br>Second step: 600 °C, 1 h, 500MPa (without Ta-tube) | *ex-situ* | HIP-S8 |
| First step: 600 °C, 11 h, ambient pressure<br>Second step: 600 °C, 1 h, 500 MPa (with Ta-tube) | *ex-situ* | HIP-S9 |
| First step: 600 °C, 11 h, 500 MPa (with Ta-tube) | *in-situ* | HIP-S10 |
| First step: 600 °C, 1 h, 500 MPa (with Ta-tube) | *in-situ* | HIP-S11 |
| First step: 600 °C, 11 h, 300 MPa (with Ta-tube) | *in-situ* | HIP-S12 |
| First step: 600 °C, 1 h, 300 MPa (with Ta-tube) | *in-situ* | HIP-S13 |

In the next step, HIP-S3 to HIP-S5 used an *in-situ* synthesis process, as discussed in the experimental part, where samples were directly placed into the HP-HST chamber without any heating process. HIP-S3 heated at 600°C for 11 h at 1 GPa with a sealed Ta-tube has the main tetragonal phase and about 26% of the hexagonal phase, which are almost identical to those of HIP-S1. HIP-S4 heated at 700°C at 1 GPa for 1 h without Ta-tube using an *in-situ* process has the same structural pattern as that of HIP-S3 and HIP-S1. It suggests that a higher heating temperature of 700°C for 1 h at 1 GPa has depicted similar behaviour for Fe(Se,Te) heated for a longer heating time at 600°C (HIP-S3) under the same growth conditions. To be more clear, the direct growth of FeSe$_{0.5}$Te$_{0.5}$ samples without a Ta-tube at 1 GPa for 11 h (*in-situ* method) has also been prepared, *i.e.*, HIP-S5 has a slightly higher hexagonal phase than that of HIP-S1 and HIP-S3. A comparative study of HIP-S1 to HIP-S5 prepared at 1 GPa reveals that an *ex-situ* FeSe$_{0.5}$Te$_{0.5}$ sample heated for 1 h (HIP-S1) has almost the same XRD pattern as that of an *in-situ* processed FeSe$_{0.5}$Te$_{0.5}$ sample heated for 11 h (HIP-S3). However, the results for HIP-S3 and HIP-S5 suggest that FeSe$_{0.5}$Te$_{0.5}$ sample sealed in a Ta tube is a more effective way to preserve the tetragonal phase than Fe(Se,Te) prepared without sealing in a Ta-tube at 1 GPa. Nevertheless, the observed hexagonal phase for all samples prepared at 1 GPa is higher than that of parent compounds, which is mentioned in Table 2.

**Table 2.** The lattice parameters and impurity phases in FeSe$_{0.5}$Te$_{0.5}$ samples under high-pressure synthesis conditions. The quantitative values of impurity phases (%) were performed using Rigaku's PDXL software and the ICDD PDF4+ 2021 standard diffraction patterns database. The hexagonal phase (H) refers to Fe$_7$Se$_8$ phase.

| Sample's code | Lattice 'a' (Å) | Lattice 'c' (Å) | FeSe$_{0.5}$Te$_{0.5}$ (%) | FeSe$_2$ (%) | Hexagonal (%) |
|---|---|---|---|---|---|
| Parent sample | 3.7950(2) | 5.9713 | 93 | - | ~6 |
| HIP-S1 | 3.8004(2) | 6.0434(5) | 70 | ~2 | ~27 |
| HIP-S2 | 3.7963(5) | 5.9578(1) | 35 | ~2-3 | ~62 |
| HIP-S3 | 3.8074(3) | 6.0838(5) | 70 | ~2-3 | ~26 |
| HIP-S4 | 3.7966(7) | 6.0075(1) | 69 | 6 | ~25 |
| HIP-S5 | 3.7971(4) | 6.0088(3) | 65 | ~2-3 | ~33 |
| HIP-S6 | 3.7977(5) | 5.9837(1) | 80 | ~3 | ~17 |
| HIP-S7 | 3.8033(9) | 6.060(2) | 30 | ~7 | ~65 |
| HIP-S8 | 3.7992(5) | 5.9843(9) | 80 | ~1-2 | ~20 |
| HIP-S9 | 3.7976(6) | 5.9679(1) | 93 | - | ~6 |
| HIP-S10 | -- | -- | 55 | ~2-3 | ~42 |
| HIP-S11 | 3.7976(5) | 5.9579(2) | 89 | ~2 | ~9 |
| HIP-S12 | ---- | --- | 26 | ~3-4 | ~70 |
| HIP-S13 | ---- | -- | 35 | ~3-4 | ~60 |

In the next step, the *in-situ* FeSe$_{0.5}$Te$_{0.5}$ sample has been heated in the HP-HTS chamber without sealing in a Ta-tube at 700 MPa for 11 h at 600°C, *i.e.*, HIP-S6. The depicted XRD pattern in Figure 2 has the dominant tetragonal phase and also around 17% of a hexagonal phase. For HIP-S7, *ex-situ* FeSe$_{0.5}$Te$_{0.5}$ was heated to 600°C and the pressure was changed from 750 MPa to 1 GPa, during the growth process. It's interesting to note that the obtained XRD pattern contains a primary hexagonal phase, as depicted in Figure 2(a)-(b), which agrees well with that of HIP-S2. HIP-S8 and HIP-S9, through an *ex-situ* process, have been prepared at 500 MPa without Ta-tube and with sealing in Ta-tube, respectively. Figure 2 depicts that both samples contain a tetragonal main phase, however, HIP-S8 has a slightly higher hexagonal phase (~20%) than that (~6%) of HIP-S9. It suggests that the *ex-situ* process of FeSe$_{0.5}$Te$_{0.5}$ sealed in a Ta-tube is a more effective way to preserve the tetragonal phase, as the same conclusion was reached from FeSe$_{0.5}$Te$_{0.5}$ samples prepared at 1 GPa. The XRD obtained for HIP-S9 is very similar to that of the parent compounds. To understand the effect of the *in-situ* process at 500 MPa, HIP-S10 and HIP-S11 have been prepared at 500 MPa for 11 h and 1 h, respectively, and sealed in Ta-tubes. The depicted XRD patterns suggest that HIP-S10 has the main phase of tetragonal but also has a huge amount of hexagonal phase (~42%), whereas the short heating time (1 h) (HIP-S11) exhibits the tetragonal phase as the main phase and has a small amount (~9%) of hexagonal phase, which is almost 5 times less than that of HIP-S10. Again, XRD patterns of HIP-S9 (*ex-situ*) and HIP-S11 (*in-situ*) suggest that FeSe$_{0.5}$Te$_{0.5}$ sealed in a Ta-tube, heated at 600°C for 1 h at 500 MPa, prepared either *ex-situ* or *in-situ*, has almost the same structural pattern as that of the parent compound. It means that FeSe$_{0.5}$Te$_{0.5}$ bulks sealed in a Ta-tube and heated for a short time (1 h) at 500 MPa are suitable for the phase formation of FeSe$_{0.5}$Te$_{0.5}$ tetragonal phase through the HP-HTS method, which is also suggested from the growth of FeSe$_{0.5}$Te$_{0.5}$ at 1 GPa. To understand the effect of low synthesis pressure (less than 500 MPa), we have prepared two more samples at 300 MPa for 11 h and 1 h, *i.e.* HIP-S12 (*in-situ*) and HIP-S13 (*in-situ*), respectively. Interestingly, both samples have the main phase of the hexagonal phase, which is enhanced with heating time, suggesting that this pressure is not suitable for the formation of the FeSe$_{0.5}$Te$_{0.5}$ tetragonal phase either for a short-term or long-term heating process. However, it also implies that a short heating process through HP-HTS is sufficient to complete the formation of the superconducting phase.

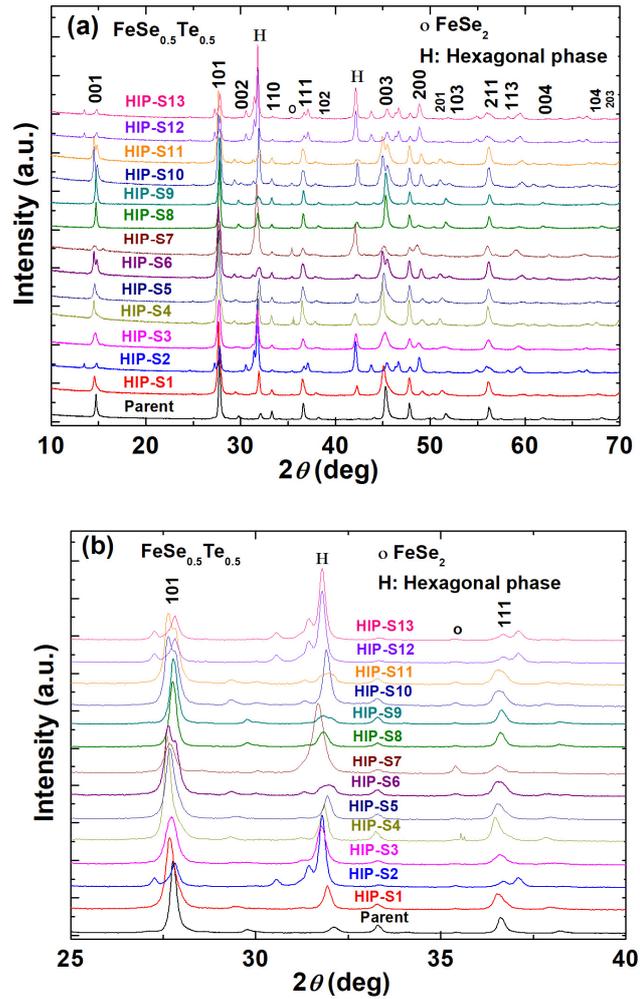

**Figure 2: (a)** Powder X-ray diffraction patterns (XRD) of various FeSe$_{0.5}$Te$_{0.5}$ prepared by the HP-HTS method. **(b)** Enlarged views of the XRD patterns, as shown in Figure *(a)*, in the 2θ range of 25 to 40 degrees for various samples. Fe$_{1.1}$Se$_{0.5}$Te$_{0.5}$ tetragonal phase was determined as the actual composition of the superconducting phase rather than the nominal composition of FeSe$_{0.5}$Te$_{0.5}$. The hexagonal phase of Fe$_7$Se$_8$ was found and has also been identified as 'H' in figures (a)-(b). The obtained lattice parameters ('*a*' and '*c*') and the foreign phases (FeSe$_2$ and Fe$_7$Se$_8$) are listed in Table 2.

These structural investigations reveal that the optimum growth conditions through HP-HTS are 500 MPa, 600°C, and 1 h with the sample sealed in a Ta-tube for the tetragonal phase formation of FeSe$_{0.5}$Te$_{0.5}$, where it works well for both *ex-situ* and *in-situ* processes. The hexagonal phase Fe$_7$Se$_8$ as an impurity phase in the samples is very sensitive to the growth pressure, sintering duration, and heating temperatures. The diffracted tetragonal peaks of the samples prepared by HP-HTS are almost at the same position as those of the parent compounds, as shown in Figure 2(b). The calculated lattice parameters based on the tetragonal phase are mentioned in Table 2, where some samples have slightly large lattice parameters, which suggests lower concentrations of Se/Te due to a large amount of impurity phases. One can observe that the presence of a large amount of impurity phase generates the large error bars in the obtained lattice parameters. The samples having a smaller amount of the impurity phases, such as HIP-S9 and HIP-S11, have almost the same lattice parameters and the same amount of hexagonal phase as the parent compounds, whereas other samples have slightly higher lattice parameters, suggesting a small deviation in Se and Te composition than that of the parent FeSe$_{0.5}$Te$_{0.5}$ composition. The parent compound has the lattice parameters ($a$ = 3.79502 Å, $c$ = 5.9713 Å) which are almost the same as the reported ones for bulk ($a$ = 3.7909 Å, $c$ = 5.9571 Å) [3, 16, 17, 7] and single crystals ($a$ = 3.815 Å, $c$ = 6.069 Å) [14, 15] of FeSe$_{0.5}$Te$_{0.5}$. In all these samples, HIP-S9 is the best sample compared to other FeSe$_{0.5}$Te$_{0.5}$ bulks through HP-HTS. Interestingly, the samples sealed in a Ta-tube or without a Ta-tube have almost the same effect on the phase formation of FeSe$_{0.5}$Te$_{0.5}$ composition. It is important to note that a non-suitable growth pressure can decrease Fe/Te/Se concentrations in FeSe$_{0.5}$Te$_{0.5}$ compositions, whereas a pressure of 500 MPa can promote the formation of a tetragonal superconducting phase.

*3.2. Elemental analysis and mapping*

The elemental analysis mapping of these samples has been performed by using the energy dispersive X-ray (EDAX) method, which allows for an understanding of the distribution of the constituent elements inside the bulk sample. The elemental mapping of the parent compound is shown in Figure 3 with the selected HIP-samples based on the XRD analysis, *i.e.,* HIP-S9, HIP-S3, and HIP-S11. Figure 3(i) confirms the homogeneous distribution of the constituent elements Fe, Se, and Te in the parent FeSe$_{0.5}$Te$_{0.5}$ compound. The mapping of HIP-S3 is depicted in Figure 3(ii). Interestingly, this sample has a non-homogeneous distribution of the constituent elements, where the large area shows Se and Fe element richness, as shown in Figure 3(ii), which suggests the formation of a hexagonal phase (Fe$_7$Se$_8$) in these samples, similar to the above discussed with XRD measurements. At a few places, Si is also observed, which could be possible during the opening of the bulk samples after the first step of the reaction (Figure 1). HIP-S9 depicts an almost homogeneous distribution of Fe, Se, and Te elements, which looks almost similar to the parent compounds and suggests the homogeneity of these constituent elements, as shown in Figure 3(iii). This analysis is consistent with the XRD data analysis. Figure 3(iv) shows the elemental mapping for HIP-S11, where the distribution of constituent elements Fe, Se, and Te is nearly homogeneous. It appears that some areas are Se-rich, which may be related to the existence of hexagonal phases as mentioned above with XRD analysis. From this data, the molar ratio of the parent compound is found to be 1.0:0.49:0.50, which is almost the same as that observed for HIP-S11 and HIP-S9, whereas in the case of HIP-S3, the observed ratio is 1.0:0.53:0.48. This ratio is more deviated in the case of other samples due to the presence of significant amounts of hexagonal phase and FeSe$_2$.

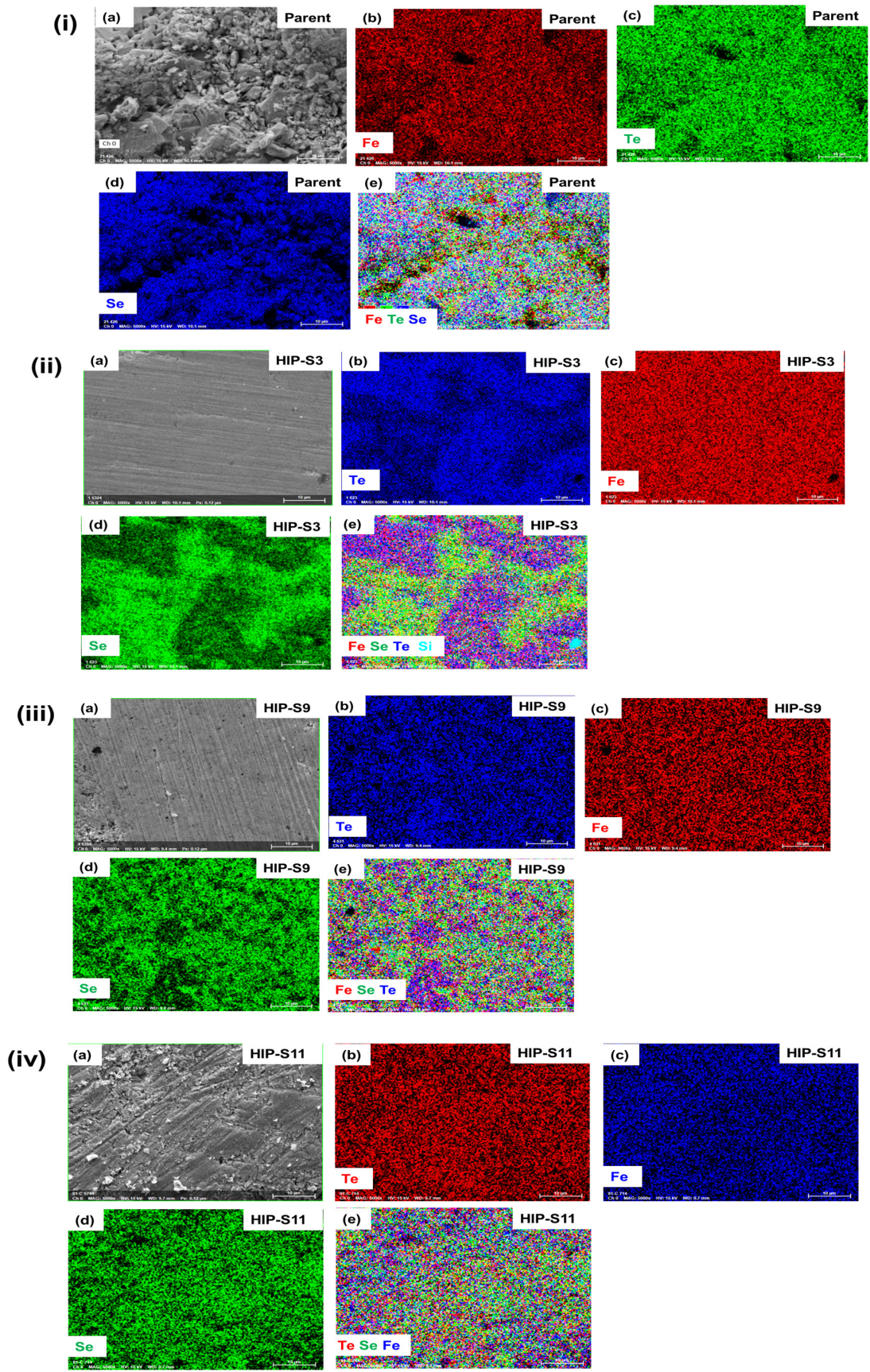

**Figure 3:** Mapping for the constituent elements of **(i)** the parent compound, **(ii)** HIP-S3, **(iii)** HIP-S9, and **(iv)** HIP-S11. Figure (a) of each image shows the area used for the mapping; Figures (b)–(d) depict the individual elemental mapping such as Fe, Se, and Te. The mapping of all elements is included in Figure (e).

*3.3. Microstructural analysis*

To comprehend the microstructural analysis of these samples, we have polished bulk samples by using several grades of micron paper, from low grade to extremely fine grade, inside the glove box to reduce any contamination of the samples due to moisture and air. Backscattered scanning electron microscopy (BSE-SEM) has been collected at different magnifications for all these samples to reveal chemical contrast. Figure 4 depicts BSE images of HIP-S3, HIP-S9, and HIP-S11 from low magnification to high magnification with the parent FeSe$_{0.5}$Te$_{0.5}$ compound. These images contain light gray, white, and black contrasts corresponding to FeSe$_{0.5}$Te$_{0.5}$, hexagonal phase Fe$_7$Se$_8$, and pores, respectively. Figure 4(a)-(c) depicts the fairly homogeneous microstructure of the parent compound, where light gray and black contrasts are observed. Many small and large pores are also observed. It appears that many grains are not well connected due to the presence of micro- or nanopores between them. However, we have not observed any impurity phases. The high-pressure synthesis method has improved grain connectivity and sample density, as shown in Figure 4. HIP-S3 bulk sample was not sealed in a Ta-tube and had an applied growth pressure of 1 GPa. Interestingly, we have observed an inhomogeneous microstructure due to the presence of white contrast related to the impurity phase of Fe$_7$Se$_8$. Because of this, the microstructure appears more compact and has fewer pores, as seen in black contrast, than the parent compounds. The observed impurity phase appeared at random, on grain boundaries, and within grains. The pores and impurity phases lead to poorer connections between the corresponding superconducting grains [36, 37]. Figure 4(g)-(i) shows the BSE images for HIP-S9, which have homogeneous gray color in the whole area and no impurity phase, similar to the observed mapping in Figure 3(iii). Interestingly, this sample is very compact, which might be due to the reduction of pores in between grains due to high-pressure synthesis effects, and we have observed a fairly homogeneous microstructure without any impurity phases, as demonstrated in a similar analysis of XRD data. In a few locations, we have observed nanopores as depicted in Figure 4(h) at a large scale. The microstructural images of HIP-S11 are depicted in Figure 4(j)-(l), which is prepared through an *in-situ* process at 500 MPa and placed into the HP-HTS chamber for 1 h. These images imply that the sample is not very compact, but in a few places, we have observed the hexagonal phase (Figure 4(k)). Pores do exist in between grains, which suggests that grains are not well connected. However, it appears that this sample has a slightly better density and grain connection than the parent compound. All these samples have disk-shaped grains. The white phase is observed in many other samples (not shown here), but one point is noticeable: the number of pores is reduced dramatically due to high-pressure growth, especially in the case of the sealed sample in the Ta-tube. The increased impurity phase that is sandwiched between FeSe$_{0.5}$Te$_{0.5}$ grains often considerably reduces grain-to-grain connections and creates a strong barrier to intergranular supercurrent routes. We did not observe any mico-cracks in any samples, as reported for other iron-based superconductors at grain boundaries and within grains [38, 36, 37]. The sample density of various samples has been calculated by assuming the pure phase of FeSe$_{0.5}$Te$_{0.5}$ for our various samples. Since the reported theoretical density of FeSe$_{0.5}$Te$_{0.5}$ is 6.99 g/cm$^3$. On this basis, the obtained density is around 51%, 70%, 44.3 %, and 61% for the parent FeSe$_{0.5}$Te$_{0.5}$, HIP-S9, HIP-S3, and HIP-S11, respectively. It indicates that the high-pressure growth of the sample sealed in a Ta-tube (HIP-S9 and HIP-S11) improved the sample density. Figure 4 clearly demonstrated that high gas pressure synthesis improves the grain connectivity and sample density, and decreases the pore size and number of pores. However, the synthesis pressures other than 500 MPa reduce the phase purity and cleanness of grain boundaries and increase the impurity phases of Fe$_7$Se$_8$ and FeSe$_2$. Pore sizes are increased when the sample is not sealed in a Ta-tube due to the gas pressure effects; however, the sample sealed in a Ta-tube has a very small number of pores compared to the parent compound or the prepared bulks without a Ta-tube. Non-superconducting phases existing at the grain boundaries of bulk FeSe$_{0.5}$Te$_{0.5}$ act as an obstacle for the transport properties and reduce the superconducting properties, as also reported for other iron-based superconductors [39, 2, 40]. As a result, our analysis suggests that high gas pressure synthesis works effectively to increase material density and also improve grain size and connectivity, but during the synthesis process, the bulks must be sealed into Ta-tubes.

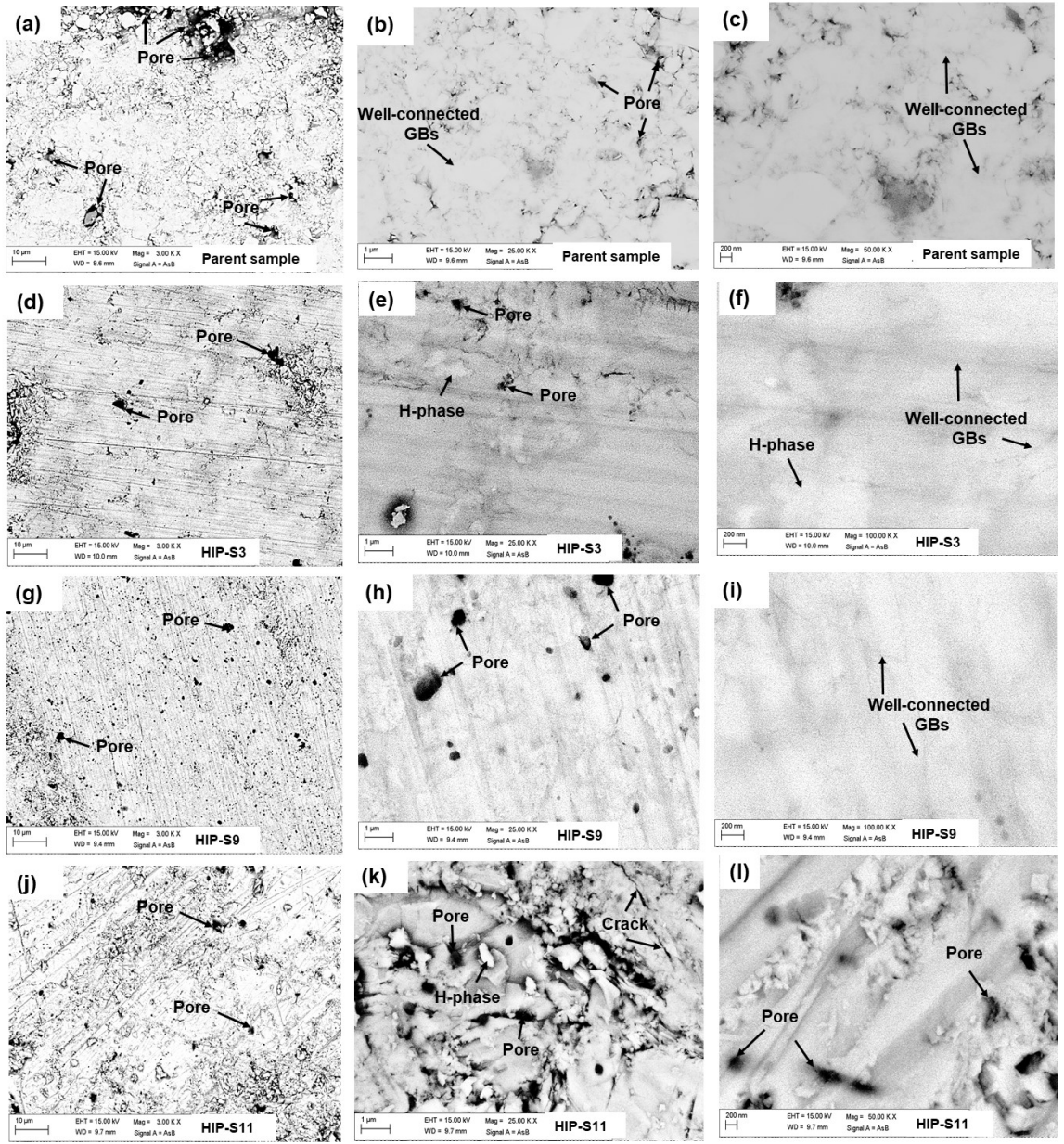

**Figure 4:** Microstructural images (Back-scattered (BSE); AsB) of **(a)-(c)** the parent compound, **(d)-(f)** HIP-S3, **(g)-(i)** HIP-S9, and **(j)-(l)** HIP-S11. Bright contrast, light gray, and black contrast correspond to the phases of hexagonal $Fe_7Se_8$ (H), $FeSe_{0.5}Te_{0.5}$, and pores, respectively.

*3.4. Transport properties*

The variation of resistivity ($\rho$) with temperature is depicted in Figure 5(a) in the full temperature range from 7 to 300 K for all $FeSe_{0.5}Te_{0.5}$ prepared by HP-HTS at 1 GPa and the parent compounds in zero magnetic field. The low-temperature behaviour of the resistivity ($\rho$), as a function of temperature from 6 K to 18 K is illustrated in Figure 5(b) for these samples. Due to the structural phase transition, the parent $FeSe_{0.5}Te_{0.5}$ exhibits a large anomaly in resistivity at a temperature below ~110 K [3, 22]. The parent compound shows a transition temperature of around 14.8 K with a transition width ($\Delta T$) of 3.7 K, as clear from Figure 5(b). HIP-S1 has a slightly higher hexagonal phase of $Fe_7Se_8$, which is a metal in the temperature range from 300 to 100 K, and below 100 K, a metal-to-insulator transition is observed [41]. Hence, the room temperature resistivity of HIP-S1 is slightly lower than that of the parent compound, but below 150 K, the resistivity enhances rapidly due to the hexagonal phase and depicts a structural transition similar to that of the parent compound. The superconducting onset transition is observed at 14 K, but the zero resistivity is not observed, suggesting the presence of a non-superconducting phase, as shown in Figure 5(b). HIP-S2 has the main hexagonal phase and almost half the resistivity of the parent compound. This normal-state resistivity is depicted as the metal-insulator transition up to the superconducting transition, which is similar to the behavior of the $Fe_7Se_8$ phase [41]. The superconducting transition of HIP-S2 starts around 14 K, but the transition is very broad and does not reach zero resistivity. The *in-situ* processed HIP-S3 has been reacted at 1 GPa and has a hexagonal phase of ~26% (Table 2). Interestingly, this sample has much lower resistivity than that of the parent compound, and the normal state resistivity depicts the metal-to-insulator transition due to the hexagonal phase [41]. Its onset superconducting transition and zero resistivity are observed at 13.9 K and 8.5 K, respectively. HIP-S4 exhibits a similar behaviour of the normal state resistivity and a comparable superconducting transition as that of HIP-S3, but has a slightly higher resistivity value than HIP-S3. This might be possible due to the presence of another phase of $FeSe_2$. The normal state resistivity is further increased for HIP-S5 due to the enhancement of the foreign phase (Table 2). Although its behaviour and superconducting transition are nearly identical to those of HIP-S4. In the case of HIP-S6, it has a lower normal state resistivity and its behaviour is similar to that of the parent compound, whereas the superconducting transition has a $T_c^{onset}$ of 11.7 K but no zero resistivity is obtained. HIP-7 is prepared under the changing pressure from 750 MPa to 1 GPa during the growth and has a main hexagonal phase. Its normal-state resistivity has depicted a similar behaviour as reported for $Fe_7Se_8$, *i.e.*, metal-to-insulator transition [41]. The normal resistivity value is increased, and the onset of the superconducting transition is observed at 12.4 K, but the zero resistivity is not reached. In all these samples prepared at 1 GPa and 700 MPa, only *in-situ* processed HIP-S3 has a clear superconducting transition with a large transition width.

Further decreasing the synthesis pressure to 500 MPa, the resistivity behaviour of these prepared samples is depicted in Figures 5(c) and 5(d). *Ex-situ* processed HIP-S8 and HIP-S9 are prepared at 500 MPa with Ta-tube and without Ta-tube, respectively. The variation of the normal resistivity is almost the same for these two samples and similar to that of the parent compound. The normal resistivity value is much lower than that of the parent compound due to the high-pressure growth effect and the compactness of the bulk sample. HIP-S8 and HIP-S9 depict the onset $T_c$ of 12.6 K and 16.2 K, whereas zero resistivity is observed at 9 K and 12.4, respectively. It implies that through HP-HTS, the preparation of $FeSe_{0.5}Te_{0.5}$ sealed inside a Ta-tube is more effective than that without a Ta-tube. To be more precise, *in-situ* processing of $FeSe_{0.5}Te_{0.5}$ at 500 MPa has been performed as HIP-S10 and HIP-S11 samples. The resistivity of HIP-S10 has a lower normal state resistivity than that of the parent compound, but its behaviour is influenced by $Fe_7Se_8$, which is presented in significant amounts, as observed from the XRD analysis. Its onset $T_c$ is around 11.9 K although zero resistivity is not reached. The observed resistivity for HIP-S11 is depicted in Figures 5(c) and (d), where the structural transition is very similar to the parent compound and also has a smaller normal state resistivity value compared to the parent

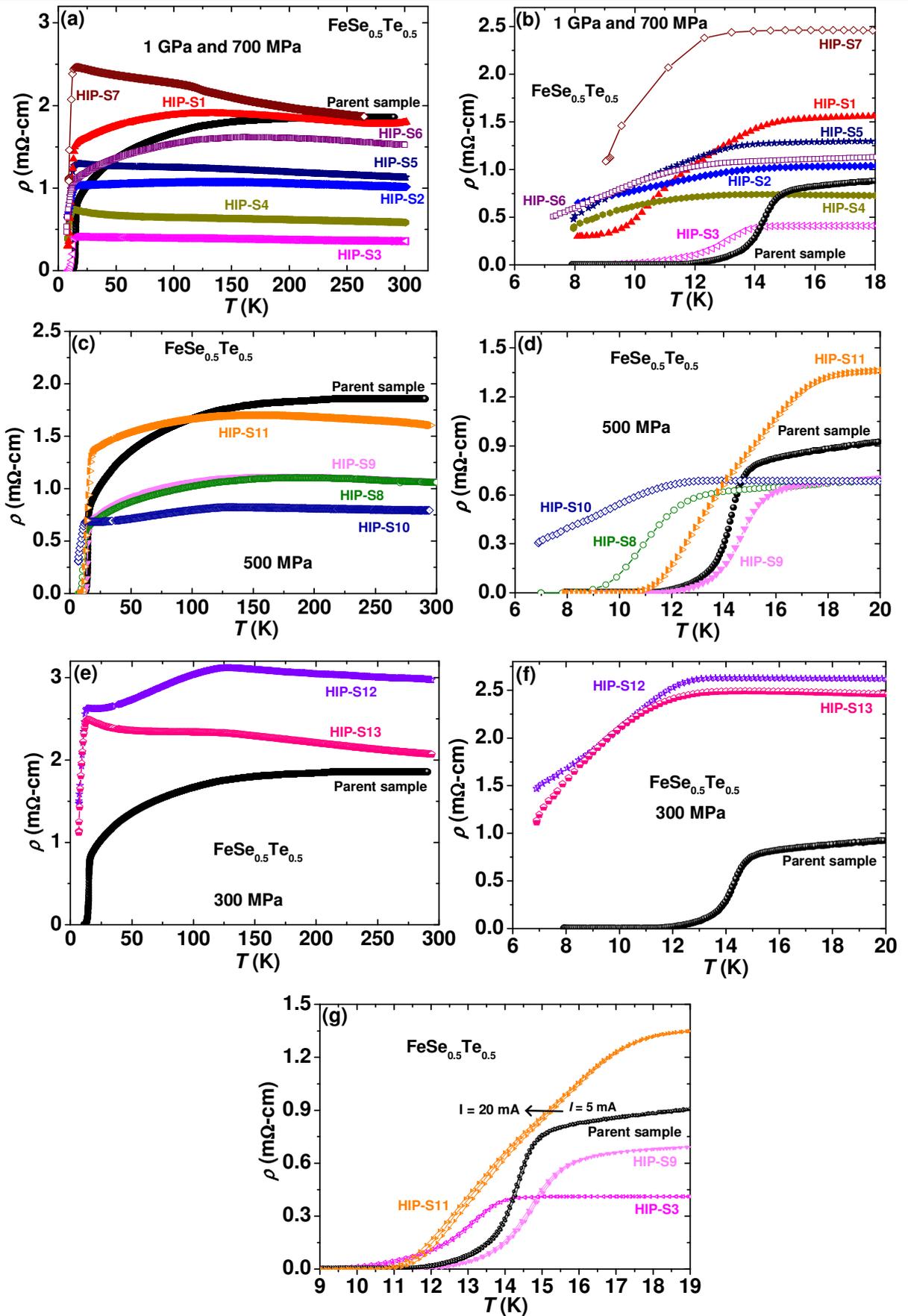

**Figure 5:** The resistivity (ρ) behaviour with the temperature parameter up to room temperature for various FeSe$_{0.5}$Te$_{0.5}$ samples prepared through the HP-HTS method at **(a)** 1 GPa and 700 MPa, **(c)** 500 MPa, and **(e)** 300 MPa. Low-temperature resistivity behaviours of various samples for the applied growth pressure **(b)** 1 GPa **(d)** 500 MPa **(f)** 300 MPa. **(g)** The temperature dependence of low-

temperature resistivity for various currents $I$ = 5, 10, and 20 mA for the parent, HIP-S3, HIP-S9, and HIP-S11.

compound before the structural transition, and after this transition, it is slightly higher before going to the superconducting transition. This sample has an onset $T_c$ of 17.3 K and an offset $T_c$ of 10.7 K. Interestingly, this onset transition is the highest in all these samples and ~2 K higher than that of the parent compound, but it is a slightly wider transition. It suggests that this $T_c$ enhancement could be due to the high concentration of actual Te/Se inside the sample. Figures 5(e) and 5(f) show the temperature variation of resistivity for the samples prepared at 300 MPa. HIP-S12 and HIP-S13 have much higher resistivity values in the whole temperature range than those of the parent compound and other bulks, which suggests a huge amount of the impurity phase. It agrees well with the XRD results, as discussed above. Both samples have an onset $T_c$ of ~12 K, but the zero resistivity transition is not observed up to 7 K, which could be due to the main phase of the $Fe_7Se_8$ phase and is well in agreement with the normal resistivity behaviour of HIP-12 and HIP-13.

These resistivity measurements suggest that high-pressure synthesis at 500 MPa for a short heating time of 1 h is the best condition compared to other growth pressures applied through the HP-HTS method. In the *ex-situ* process, the samples sealed in a Ta-tube (HIP-S9) have a 1 K higher $T_c^{onset}$ value with a slightly sharper transition width (~3.1 K) than that of the parent compound, whereas in the *in-situ* process *i.e.*, the bulks sealed in a Ta-tube (HIP-S11) have enhanced the transition temperature by ~2 K and have a broader transition width. However, *in-situ* processed $FeSe_{0.5}Te_{0.5}$ without sealing into a Ta-tube (HIP-S8) has a slightly higher hexagonal phase than that of the parent compound and HIP-S9, which might be a reason for the low superconducting transition. These results are in good agreement with those discussed above with XRD and microstructure analysis.

To understand the intergrain and intragrain connections, we have measured the temperature dependence of resistivity behaviour under various applied currents [42, 36, 37]. Generally, the onset of the superconducting transition ($T_c^{onset}$) reflects the individual grain effects, and the offset of the transition ($T_c^{offset}$) is related to the grain connections, *i.e.*, intergrain connections [39, 43, 40]. Based on the aforementioned discussion, we have measured the low-temperature resistivity under various currents, I = 5, 10, and 20 mA, which are depicted in Figure 5(g) for HIP-S3, HIP-S9, HIP-S11, and the parent compound. The bulk samples of HIP-S3 have a lower $T_c$ and a broad transition with the applied current up to 20 mA, where the offset transition is more sensitive and suggests a weak grain connection compared to the parent compound. HIP-S9 has a higher onset $T_c$ of 16.2 K and a sharp transition broadening compared to that of the parent compound. The offset transition has almost no transition broadening with various currents, suggesting a good grain connection. HIP-S11 has an onset $T_c$ of ~17.2 K, and the broadening of the transition is observed at the offset transition with various applied currents, as shown in Figure 5(g), which could be due to the presence of impurity phases in between grains, as discussed from XRD patterns. These outcomes support the analysis of microstructural studies, as observed above. Compared to all these samples, HIP-S9 samples exhibit better grain connections, whereas the offset transition broadening is more clearly observed for HIP-S3 and HIP-S11.

*3.5. Magnetic property measurements*

To confirm the Meissner effect, we have measured DC magnetic susceptibility ($\chi$ = $4\pi M/H$) in both zero-field-cooled (ZFC) and field-cooled (FC) magnetization curves for four samples: the parent compound, HIP-S9, HIP-S3, and HIP-S11, in the temperature range from 5 to 22 K under the applied magnetic field of 50 Oe. The normalized magnetic susceptibility is shown for all samples in Figure 6(a) so that we can make a comparison study. The small diamagnetic signal in FC data suggests the strong pinning nature of the bulk sample and the permanent flux trapped inside it. The volume fraction of the superconducting phase was estimated from the measured magnetic susceptibility data ($\chi_{ZFC}$) by correcting the demagnetization factor, as described in [44, 45, 46], and was obtained ~92% for the parent compound, which is similar to previous reports [47]. HIP-S9, HIP-S11, and HIP-S3 have the superconducting volume fractions of ~97%, ~95% and ~61%, respectively.

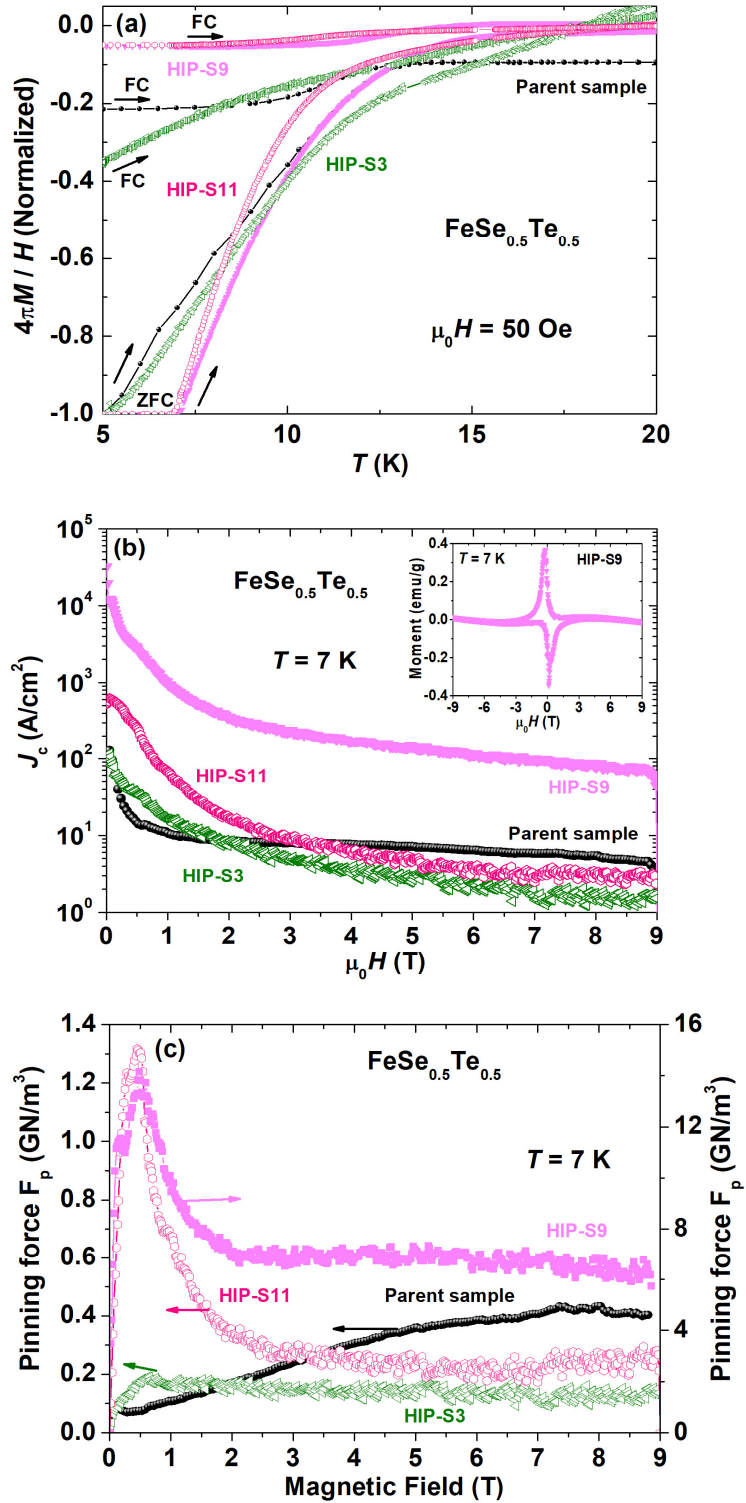

**Figure 6: (a)** The temperature dependence of the normalized magnetic susceptibility ($\chi = 4\pi M / H$) for three samples: parent compound, HIP-S3, HIP-S9, and HIP-S11 under $\mu_0 H$ = 50 Oe magnetic field in the temperature range 5-20 K through zero field-cooled (ZFC) and field-cooled (FC) regimes. **(b)** The magnetic field ($H$) dependence of critical current density ($J_c$) for the parent FeSe$_{0.5}$Te$_{0.5}$ and HIP-S3, HIP-S9, and HIP-S11 bulk samples at 7 K. The inset figure shows the magnetic hysteresis loop $M(H)$ at 7 K for HIP-S9 after the subtraction of the normal state background. **(c)** The magnetic field

dependence of the pinning force density ($F_p$) was calculated from the critical current density for the parent compound, HIP-S3, HIP-S9, and HIP-S11 at 7 K. The arrow indicates the used scale for the respective bulks.

These measurements confirm the bulk superconductivity of these materials. A superconducting transition is observed at 14 K with a diamagnetic transition in the magnetic susceptibility ($\chi$) in both the ZFC and FC situations for the parent compound. HIP-S9 shows the onset transition at 15.1 K and has a sharper transition and a high superconducting volume fraction as compared to other samples. However, HIP-S3 has a slightly lower onset $T_c$ value than that of the parent FeSe$_{0.5}$Te$_{0.5}$, however, the magnetic transition has a large broadening of the transition and a low superconducting volume fraction due to the presence of impurity phases, which was also confirmed by the resistivity measurements. The slightly lower onset $T_c$ suggests the reduction of the actual content of Te or Se from the main phase FeSe$_{0.5}$Te$_{0.5}$, as discussed above with the XRD data and the microstructural analysis.

HIP-S11 depicts a higher onset transition of 16.1 K, however, has a broader transition compared to the parent compound and HIP-S9. Interestingly, all samples depict the single-step transition, which can be explained by the intergranular properties of these bulk samples, as discussed and reported for other FBS families [39]. Fe(Se,Te) prepared by the melting route [20] exhibits a double and broad transition in the magnetization measurements, suggesting weak grain connections. In comparison to that [20, 24], HIP-S9 and HIP-S11 have sharper transitions and a higher $T_c$ value, indicating the improved sample quality by the HP-HTS method. These analyses confirm that high-pressure growth at 500 MPa, *i.e.* HIP-S9 and HIP-S11, is an effective condition for the superconducting properties of FeSe$_{0.5}$Te$_{0.5}$, similar to the conclusion of the above-discussed measurements. The reduction of $T_c$ observed for HIP-S3 could be related to changes in Te/Se concentrations and the formation of foreign phases.

We have measured magnetic hysteresis loops (*M-H*) at a constant temperature of 7 K for three HIP samples (HIP-S3, HIP-S9, and HIP-S11) and the parent compound with the rectangular-shaped sample to calculate the critical current density $J_c$. The measured magnetic loops $M(H)$ for these samples were observed under ferromagnetic effects, which is similar to previous reports based on Fe(Se,Te) samples [23, 7, 22, 21]. *M-H* loop for HIP-S9 is depicted as an inset figure 6(b) after the subtraction of the normal state magnetization, *i.e.* $M(H)$ loop at 22 K. Bean critical state model [48] is applied to obtain the critical current densities by using these magnetization hysteresis loops. Bean model estimates the critical current density ($J_c$) of a superconductor assuming that supercurrents flow with a density equal to the critical-current density $J_c(H)$, which remains a constant independent of the magnetic field ($H$). Several other critical state models were proposed by Kim *et al.* or Kim-Anderson *et al.* [49], to extend the critical state model for various applications and to incorporate the field dependence of critical current density. However, Bean model is the most common and widely used method to estimate the critical current density from *M-H* curves, in which the value of $J_c(T, H)$ does not vary much with the magnetic field for low temperature isotherms. Bean's model gives a good estimation of the critical current density and provides a simple, intuitive framework in which data could be analyzed, despite the need for approximations when applying the model to samples of finite dimensions. Furthermore, this model is applied to the critical current analysis of all other iron-based superconductors [44, 2, 4]. Hence, the calculation of the critical current density $J_c$ for our samples was performed using the Bean formula $J_c = 20\Delta m/Va(1- a/3b)$, where $\Delta m$ is the hysteresis loop width, $V$ is the volume of the sample, and $a$ and $b$ are the lengths of the shorter and longer edges, respectively. Figure 6(b) depicts the magnetic field dependence of the critical current density ($J_c$) up to 9 T at 7 K for the parent compound with three samples: HIP-S3, HIP-S9, and HIP-S11. Interestingly, the calculated $J_c$ of HIP-S3 has a field dependence very similar to that of the parent compound. However, this sample has a slightly better $J_c$ value in the low magnetic field (≤ 2 T) region, whereas the $J_c$ value is lower in the high magnetic field compared to that of the low magnetic field. It could be due to the reduced density and grain connections, especially in the high magnetic field, which are clearly observed in the microstructural analysis and resistivity studies. The obtained $J_c$ value of HIP-S9 (*ex-situ*) has depicted a huge improvement in the critical current properties in the whole magnetic field range, which is almost two orders of magnitude higher than the parent compound and HIP-S3 (*in-situ*). In the case of HIP-S11 (*in-situ*), the $J_c$ value is improved by one order magnitude in the low field region below the magnetic field of 3 T; however, in the high field, the $J_c$ value is almost the same as that of the parent compound. The enhancement of $J_c$ could be explained by the introduction of flux pinning centres by high-pressure synthesis growth [26]. The enhanced sample density and improved grain connections can be another factor in the critical current improvement, which is observed from the microstructural and resistivity analyses [39]. For example, HIP-S9 has 20% higher density and better grain connectivity than the parent compound, whereas lower density is observed for HIP-S11 and HIP-S3. It suggests that high-pressure synthesis at 500 MPa and a sample sealed into a Ta-tube can work well to enhance the magnetic field response of $J_c$ for bulk FeSe$_{0.5}$Te$_{0.5}$, which is comparable to the reported elevation of $J_c$ values for Fe(Se,Te) samples [20, 50]. Chemical addition generally plays an important role in improving the intergrain connections and enhancing the critical current properties. Many studies have been reported for Fe(Se,Te) with various chemical additions [50, 22, 21], and interestingly, the $J_c$ value obtained in our study is higher than the reported $J_c$ for the metal-added FeSe$_{0.5}$Te$_{0.5}$. Hence, high-pressure growth conditions work more effectively to improve the intergranular current than other methods, such as metal additions [22]. On the other hand, FeSe$_{0.5}$Te$_{0.5}$ bulks prepared by the melting method [20, 24] and very long annealing procedures such as 740 h depicted a critical current density almost similar to our findings. However, such a long annealing always reduces the actual contents of Se and Te from FeSe$_{0.5}$Te$_{0.5}$ composition and results a lower superconducting transition, as observed in previous reports [20, 24, 3].

The enhancement of critical current density is directly related to the pinning force. To analyze the pinning force behaviours for these samples, we have calculated the variation of vortex pinning force density $F_p$ with the applied magnetic field dependence by using the relation $F_p = \mu_0 H \times J_c$, as illustrated in Figure 6(c), where the critical current density $J_c$ is calculated using Bean model as shown in Figure 6(b) for these samples. The pinning force of the parent compound increases with the magnetic field and reaches its maximum at 8 T, and the obtained $F_p$ values of the parent compounds are almost the same as those reported (0.1–1 GN/m$^3$) in previous studies based on polycrystalline Fe(Se,Te) samples [51, 52, 22]. Interestingly, all these samples prepared by HP-HTS have completely different behaviour than the parent compound. For the HIP-S3 sample, the pinning force increases with the magnetic field and reaches its maximum at around 0.5 T. With further increases in the magnetic field, the pinning force slightly decreases and becomes almost constant over the whole magnetic field range. However, the $F_p$ value of HIP-S3 is slightly higher up to the magnetic field of 2 T, and after that, its value is much lower than that of the parent compound. The sample prepared at 500 MPa and sealed in a Ta-tube using an *ex-situ* process, *i.e.* HIP-S9, has a huge enhancement of the pinning force compared to other samples; however, its behaviour is very similar to that of HIP-S3. The maximum pinning force is observed at 0.5 T; however, the field independence behaviour of $F_p$ is observed in the magnetic field range 2–9 T, which is in good agreement with the $J_c$ behaviours as depicted in Figure 6(b). HIP-S11, *i.e.*, the sealed sample in a Ta-tube and prepared at 500 MPa using an *in-situ* process, has the same behaviour as HIP-S3 and HIP-S9, and the observed pinning force has better performance with the applied magnetic field (≤3T) compared with the parent compound and HIP-S3. This different behaviour of pinning force for HIP-S3, HIP-S9, and HIP-S11 compared to the parent FeSe$_{0.5}$Te$_{0.5}$ compound is most likely caused by high-pressure synthesis effects, which warrants further investigation to understand how synthesis pressure can influence the vortex pinning mechanisms [53] in FeSe$_{0.5}$Te$_{0.5}$ compounds. One can note that the $F_p$ values of HIP-S9 are similar to those reported for Fe(Se,Te) single crystals (10-1000 GN/m$^3$) [52]. Our analysis of $F_p$ behaviour leads us to the conclusion that improving the microstructure, material density, and the appropriate pinning centres may contribute to improving the pinning properties and the enhancement of the critical current behaviours as reported for FBS [26, 42] and MgB$_2$ superconductors [54]. This pinning force analysis well agrees with the above-mentioned structural and microstructural analyses.

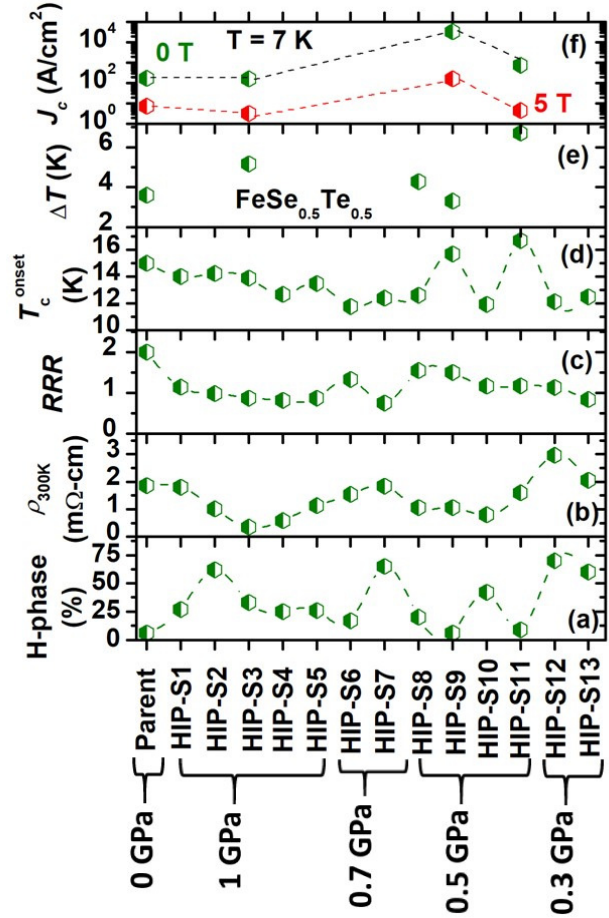

**Figure 7:** The variation of **(a)** hexagonal (H) phase (%) calculated from XRD patterns, **(b)** room temperature resistivity ($\rho_{300K}$), **(c)** residual resistivity ratio $RRR$ ($\rho_{300K} / \rho_{20K}$), **(d)** transition temperature ($T_c$) **(e)** transition width ($\Delta T$) and **(f)** the critical current density $J_c$ for 0 T and 5 T at 7 K for various samples concerning the parent FeSe$_{0.5}$Te$_{0.5}$.

## 4. Discussion

We have summarized the main findings of our study in Figure 7(a)-(f), where the variation of the hexagonal (*H*) phase, the room temperature resistivity ($\rho_{300K}$), the residual resistivity ratio ($RRR = \rho_{300K}/\rho_{20K}$), the onset transition temperature $T_c^{onset}$, the transition width ($\Delta T = T_c^{onset} - T_c^{offset}$), and the critical current density ($J_c$) for 0 T and 5 T at 7 K concerning various samples are shown. Figure 7(a) shows that the parent compound has around 6% of the hexagonal phase, as also reported by previous studies [18, 23, 7]. The *ex-situ* process of FeSe$_{0.5}$Te$_{0.5}$ sealed in a Ta-tube at 1 GPa for 1 h, *i.e.*, HIP-S1, has enhanced the hexagonal phase. Further increased sintering time up to 11 h, *i.e.* HIP-S2 has a main phase of hexagonal phase. *In-situ* processing, *i.e.* direct heating of FeSe$_{0.5}$Te$_{0.5}$ either sealed in a Ta-tube or without a Ta-tube at 1 GPa (HIP-S3, HIP-S4, HIP-S5), leads to almost the same amount of hexagonal phase, which is lower than that for HIP-S1 and HIP-S2. This suggests that the tetragonal phase formation at 1 GPa is not so effective either when FeSe$_{0.5}$Te$_{0.5}$ sample is sealed in a Ta-tube or without sealing in a Ta-tube *i.e.*, this pressure is not supportive for the tetragonal phase formation. The sample prepared at 700 MPa using *ex-situ* also has a hexagonal phase, which is enhanced very rapidly and became the main phase during changing the pressure from 750 MPa to 1 GPa (HIP-S7). The growth pressure of 500 MPa using *ex-situ* samples sealed into a Ta-tube (HIP-S9) and without a Ta-tube (HIP-S8) has given a low amount of hexagonal phase and reached the minimum for HIP-S9. In the case of an *in-situ* process, long-time heating with Ta-tube (HIP-S10) has enhanced the hexagonal phase more than short-time heating with Ta-tube (HIP-S11). It suggests that the synthesis pressure of 500 MPa with a Ta-tube for a short reaction time is favourable for FeSe$_{0.5}$Te$_{0.5}$ phase formation. At a further lower synthesis pressure, *i.e.* 300 MPa, the hexagonal phase appeared as the main phase formation either for a long-term or short-term heating process.

The room temperature resistivity is presented in Figure 7(b). One can note that the resistivity depends on the grain connectivity and the presence of the foreign phase inside the grains or at the grain boundaries. The variation of $\rho_{300K}$ for all prepared samples at 1 GPa and 700 MPa suggests that $\rho_{300K}$ has a lower value for the samples having a large amount of hexagonal phase, such as HIP-S2. It might be due to the metallic nature of the Fe$_7$Se$_8$ phase [41]. As the hexagonal phase decreased, $\rho_{300K}$ started to increase, such as for HIP-S4 to HIP-S6. HIP-S7 has a slightly high $\rho_{300K}$ and even has the main phase of the hexagonal phase. The samples prepared at 500 MPa have also followed the same behaviour as the variation of $\rho_{300K}$ with the hexagonal phase, as discussed. In the case of 300 MPa, the samples have the highest value of $\rho_{300K}$ compared to all other samples, which could be due to the presence of foreign phases and poor grain connections.

Figure 7(c) demonstrates that the residual resistivity ratio ($RRR = \rho_{300K}/\rho_{20K}$) value is reduced for all HIP samples prepared at various pressures and reaches around 1-1.5, even though some samples have the main phase of hexagonal phase. It could be possible to have good compensation between the impurity phase, grain connectivity, and chemical homogeneity due to the high-pressure synthesis effect. The variation of onset $T_c$ is shown in Figure 7(d). It suggests that the presence of the hexagonal phase reduces the $T_c^{onset}$ and reaches the maxima for the sample having a low amount of foreign phases. HIP-S9 and HIP-S11 have a hexagonal phase that is almost the same as that of the parent compound and depicts the highest transition temperature up to 17.2 K. As shown in Figure 4, it is clear that the resistivity of many samples did not reach zero resistivity, which suggests a high amount of hexagonal phase, *i.e.*, non-superconducting phase. A lower $T_c$ value than that of the parent compound suggests a lower Se/Te composition in the nominal composition of FeSe$_{0.5}$Te$_{0.5}$. The transition width $\Delta T$ variation is shown in Figure 7(e) for various samples that have shown the zero resistive transition during the resistivity measurements. HIP-S3, HIP-S8, and HIP-S11 have very broad transitions, whereas HIP-S9 has a transition width of 3.1 K, which is slightly sharper than that of the parent compound ($\Delta T$~3.6), suggesting better intergrain connections. Figure 7(f) shows the variation of $J_c$ at 7 K for 0 T and 5 T magnetic fields for three samples and the parent compounds. Interestingly, HIP-S9 has an almost two orders of magnitude higher $J_c$ value than that of the parent compound, whereas HIP-S3 and HIP-S11 have almost the same order of magnitude as the parent compound, but at a higher magnetic field, this value has reduced rapidly, probably due to weak grain connections and pinning.

This analysis suggests that the *ex-situ* sample, *i.e.*, HIP-S9, has a low amount of hexagonal phase, half the resistivity of the parent compound, $RRR$ of 1.5, and a slightly (~1 K) higher transition temperature, the shaper transition ($\Delta T$ = 3.1 K). Interestingly, this sample has an almost two-times higher critical current density than that of the parent compound in the whole magnetic field region. This suggests that high-pressure growth of an *ex-situ* processed FeSe$_{0.5}$Te$_{0.5}$ sample at 500 MPa and 600°C, sealed in a Ta-tube provides better conditions to improve the superconducting transition and a huge improvement in the critical current density. *In-situ* processed FeSe$_{0.5}$Te$_{0.5}$ samples prepared at 500 MPa for 1 h and sealed into Ta-tubes have the highest transition of 17.2 K and one order of magnitude higher $J_c$ value than the parent compound but are smaller than *ex-situ* processed HIP-S9. Rest samples have low performance with the superconducting properties due to the formation of the hexagonal phase, which seems to be more sensitive to high pressure and high synthesis temperatures.

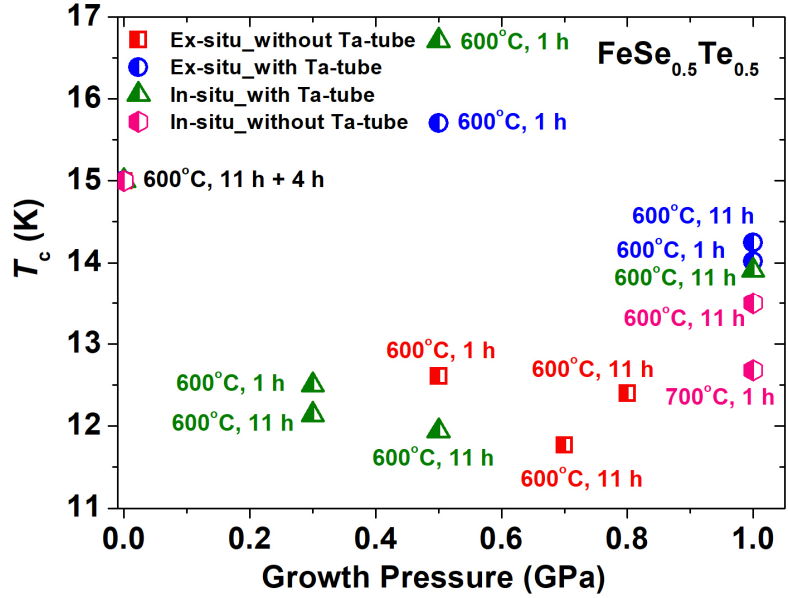

**Figure 8:** The variation of the onset transition temperature ($T_c$) of various FeSe$_{0.5}$Te$_{0.5}$ bulks concerning the applied growth pressure for various FeSe$_{0.5}$Te$_{0.5}$ samples. The bulks prepared through the changing pressure from 750 MPa to 1 GPa (HIP-S7) are considered and depicted at 0.8 GPa in this figure.

Figure 8 shows the variation of $T_c^{onset}$ concerning the applied growth pressure, which suggests that *in-situ* and *ex-situ* processed FeSe$_{0.5}$Te$_{0.5}$ samples at 500 MPa are very effective to enhance the superconducting properties, and only a short heating time of 1 h is sufficient. In the case of CSP at the ambient pressure, the preparation of FeSe$_{0.5}$Te$_{0.5}$ (the parent compound) requires at least 11 h in the first step at 600°C and to compact the powder, regrinding, and heating process at 600°C, 4 h are also needed to complete the whole reaction and produce a good-quality parent compound [22, 21]. Our high-pressure investigation reveals that this long heating process (~15 h at ambient pressure) could be completed in just 1 h under the presence of the growth pressure of 500 MPa, along with the enhanced transition temperature and the critical current properties, as illustrated in Figures 7 and 8. In the case of *ex-situ* processed samples, it also suggests that 500 MPa and a heating time of 1 h are the most effective growth conditions for the superconducting phase formation of Fe(Se,Te) and to improve the superconducting properties, *i.e.*, HIP-S9, as shown in Figures 7 and 8. Furthermore, it is also evident that the HP-HTS approach performs effectively when the initial bulk sample is sealed within a Ta-tube rather than the open sample (without sealing in a Ta-tube) inside our HP-HTS chamber. *In-situ* processing is sufficient for the formation of FeSe$_{0.5}$Te$_{0.5}$ phase but is not very effective for the grain connectivity process and is also not very safe, whereas *ex-situ* processing under the same growth conditions as the *in-situ* process, works well for the intergrain connections as well as the tetragonal phase formation and is safer and easier to handle than the *in-situ* process, as reported for MgB$_2$ [55]. Additionally, more devoted research based on high-pressure synthesis would be required to further optimize the superconducting properties in bulk FeSe$_{0.5}$Te$_{0.5}$ by improving sample quality and grain connection, which will also help to understand the growth pressure effect.

## 5. Conclusions

In summary, we have prepared bulk FeSe$_{0.5}$Te$_{0.5}$ through the high gas pressure synthesis technique employing *ex-situ* and *in-situ* procedures and studied its superconducting properties with the parent FeSe$_{0.5}$Te$_{0.5}$ compound produced by CSP at ambient pressure.

Structural analysis suggests that high synthesis pressure for FeSe$_{0.5}$Te$_{0.5}$ phase promotes the hexagonal phase, but this hexagonal phase reached its lowest level for the samples prepared at 500 MPa for 1 h with sealing into a Ta-tube. Microstructural analysis confirms that FeSe$_{0.5}$Te$_{0.5}$ sample sealed into a Ta-tube is more effective for the enhancement of sample density and the improvement of grain connectivity, although the existence of a hexagonal phase depending on the pressure synthesis conditions reduces the superconducting grain connections. Especially, the bulk samples prepared without sealing into Ta-tubes diminish the grain connections due to the presence of large pores caused by the high gas pressure passing through micro/nanopores. Resistivity and magnetic measurements approve the enhancement of the transition temperature by 2 K for FeSe$_{0.5}$Te$_{0.5}$ prepared at 500 MPa. Due to a varied Se/Te concentration or a large amount of impurity phases, other samples had a lower transition temperature than the parent compound. Our studies confirm that the optimum high-pressure growth conditions are 500 MPa, 600°C, a heating time of 1 h, and the sample sealed in a Ta-tube, where the *in-situ* process is sufficient for the development of the tetragonal phase formation, and the *ex-situ* process contributes to the improvement of the intergrain connections and also promotes the formation of the superconducting phase. Critical current density and pinning force analysis corroborate that FeSe$_{0.5}$Te$_{0.5}$ prepared under these optimum conditions has almost two orders of magnitude greater enhancement of $J_c$ than the parent compound due to the improved grain connections, the enhanced sample density, and the presence of additional pinning centres. We believe that this high-pressure synthesis method will enable the further exploration of Fe(Se, Te) and other FBS materials to improve the sample quality, achieve additional improvements in their superconducting properties, and further the development of their magnetic applications, particularly for superconducting wires and tapes.

**Funding:** This research was funded by NATIONAL SCIENCE CENTRE (NCN), POLAND, grant number "2021/42/E/ST5/00262" (SONATA-BIS 11). SJS acknowledges financial support from National Science Centre (NCN), Poland through research Project number: 2021/42/E/ST5/00262.

**References**

[1] Kamihara, Y.; Watanabe, T.; Hirano, M.; Hosono, H., "Iron-Based Layered Superconductor La[O$_{1-x}$F$_x$]FeAs ($x$ = 0.05–0.12) with $T_c$ = 26 K," *J. Am. Chem. Soc.*, vol. 130, p. 3296, 2008.

[2] Singh, S. J.; Sturza, M., "Bulk and Single Crystal Growth Progress of Iron-Based Superconductors (FBS): 1111 and 1144," *Crystals*, vol. 12, p. 20, 2022.

[3] Takano, Y.; Mizuguchi, Y., "Review of Fe Chalcogenides as the Simplest Fe-Based Superconductor," *J. Phys. Soc. Jpn.*, vol. 79, p. 102001, 2010.

[4] Hosono, H.; Yamamoto, A.; Hiramatsu, H.; and Ma, Y., "Recent advances in iron-based superconductors toward applications," *Materials Today*, vol. 21, no. 3, pp. 278-302, 2018.

[5] Si, Q.; Yu, R.; and Abrahams, E., "High-temperature superconductivity in iron pnictides and chalcogenides," *Nature Reviews Materials*, vol. 1, p. 16017, 2016.

[6] Shimoyama J., "Potentials of iron-based superconductors for practical future materials," *Supercond. Sci. Technol.*, vol. 044002, p. 27, 2014.

[7] Mizuguchi, Y.; Tomioka, F.; Tsuda, S.; Yamaguchi, T.; Takano, Y., "Substitution Effects on FeSe Superconductor," *J. Phys. Soc. Jpn.*, vol. 78, p. 074712, 2009.

[8] Sales, B. C.; Sefat, A. S.; McGuire, M. A.; Jin, R. Y.; Mandrus, D.; and Mozharivskyj, Y., "Bulk superconductivity at 14 K in single crystals of Fe$_{1+y}$Te$_x$Se$_{1-x}$," *Phys. Rev. B*, vol. 79, p. 094521, 2009.

[9] Predel, B., "Fe-Se (Iron-Selenium) Dy-Er–Fr-Mo (Berlin:," *Springer-Verlag Berlin Heidelberg,* no. DOI: 10.1007/10474837_1339, 1982.

[10] Okamoto, H., "The Fe–Se (iron-selenium) system," *J. Phase Equilibria,* vol. 12, no. doi: https://doi.org/10.1007/BF02649932, p. 383, 1991.

[11] Yeh, K.-W.; Huang, T.-W.; Huang, Y.-L.; Chen, T.-K.; Hsu, F.-C.; Wu, P.M.; Lee, Y.-C.; Chu, Y.-Y.; Chen, C.-L.; Luo, J.-Y.; Yan, D.-C.; and Wu, M.-K., "Tellurium substitution effect on superconductivity of the $\alpha$-phase iron selenide," *Europhysics Letters,* vol. 84, p. 37002, 2008.

[12] Hsu, F.-C.; Luo, J.-Y.; Yeh, K.-W.; Chen, T.-K.; Huang, T.-W.; Wu, P. M.; Lee, Y.-C.; Huang, Y.-L.; Chu, Y.-Y.; Yan, D.-C.; and Wu, M.-K., "Superconductivity in the PbO-type structure $\alpha$-FeSe," *Proc. Natl. Acad. Sci. U. S. A.,* vol. 105, p. 14262, 2008.

[13] Margadonna, S.; Takabayashi, Y.; Ohishi, Y.; Mizuguchi, Y.; Takano, Y.; Kagayama, T.; Nakagawa, T.; Takata, M.; and Prassides, K., "Pressure evolution of the low-temperature crystal structure and bonding of the superconductor FeSe," *Phys. Rev. B,* vol. 80, p. 064506, 2009.

[14] Her, J. L.; Kohama, Y.; Matsuda, Y. H.; Kindo, K.; Yang, W.-H.; Chareev, D. A.; Mitrofanova, E. S.; Volkova, O. S.; Vasiliev, A. N.; and Lin, J.-Y., "Anisotropy in the upper critical field of FeSe and FeSe$_{0.33}$Te$_{0.67}$ single crystals," *Supercond. Sci. Technol.,* vol. 28, p. 045013, 2015.

[15] Viennois, R.; Giannini, E.; Marel, D. van der; Černý R., "Effect of Fe excess on structural, magnetic and superconducting properties of single-crystalline Fe$_{1+x}$Te$_{1-y}$Se$_y$," *J. Solid State Chem.,* vol. 183, p. 769, 2010.

[16] Zhang, Y.; Wang, T.; Wang, Z.; Xing, Z., "Effects of Te- and Fe-doping on the superconducting properties in Fe$_y$Se$_{1-x}$Te$_x$ thin films," *Scientific Reports,* vol. 12, p. 391, 2022.

[17] Jung, S. G.; Kang, J. H.; Park, E.; Lee, S.; Lin, J.-Y.; Chareev, D. A.; Vasiliev, A. N.; & Park, T., "Enhanced critical current density in the pressure-induced magnetic state of the high-temperature superconductor FeSe," *Scientific Reports,* vol. 5, p. 16385, 2015.

[18] Kumar, R. S.; Zhang, Y.; Sinogeikin, S.; Xiao, Y.; Kumar, S.; Chow, P.; Cornelius, A. L.; and Chen, C., "Crystal and electronic structure of FeSe at high pressure and Low temperature," *J. Phys. Chem. B,* vol. 114, p. 12597, 2010.

[19] Nouailhetas, Q.; Koblischka-Veneva, A.; Koblischka, M. R.; Naik, P.; Schäfer, F.; Ogino, H.; Motz, C.; Berger, K.; Douine, B.; Slimani, Y.; and Hannachi, E., "Magnetic phases in superconducting, polycrystalline bulk FeSe samples," *AIP Advances,* vol. 11, p. 015230, 2021.

[20] Masi, A.; Alvani, C.; Armenio, A. A.; Augieri, A.; Barba, L.; Campi, G.; Celentano, G.; Chita, G.; Fabbri, F.; Zignani, C. F.; Barbera, A. L.; Piperno, L.; Rizzo, F.; Rufoloni, A.; Silva, E.; Vannozzi, A.; and Varsano, F., "Fe(Se,Te) from melting routes: the influence of thermal processing on microstructure and superconducting properties," *Supercond. Sci. Technol.,* vol. 33, p. 084007, 2020.

[21] Singh, S.J.; Diduszko, R.; Iwanowski, P.; Cetner, T.; Wisniewski, A.; Morawski, A., "Effect of Pb addition on microstructure, transport properties, and the critical current density in a polycrystalline FeSe$_{0.5}$Te$_{0.5}$," *Applied Physics A,* vol. 128, p. 476, 2022.

[22] Manasa, M.; Azam, M.; Zajarniuk, T.; Diduszko, R.; Cetner, T.; Morawski, A.; Wiśniewski, A.; Singh, S. J., "Cometal Addition Effect on Superconducting Properties and Granular Behaviours of Polycrystalline FeSe$_{0.5}$Te$_{0.5}$," *Materials,* vol. 16, p. 2892, 2023.

[23] Shahbazi, M.; Cathey, H. E.; and Mackinnon, I. D. R., "Stoichiometry of tetragonal and hexagonal Fe$_x$Se: phase realtions," *Supercond. Sci. Technol.,* vol. 33, p. 075003, 2020.

[24] Li, X.; Shi, X.; Wang, J.; Zhang, Y.; Zhuang, J.; Yuan, F.; Shi, Z., "Synthesis of high-quality FeSe$_{0.5}$Te$_{0.5}$ polycrystal using an easy one-step technique," *Journal of Alloys and Compounds,* vol. 644, pp. 523-527, 2015.

[25] Ma, Y.; Zhang, X.; Nishijima, G.; Watanabe, K.; Awaji, S.; Bai, X., "Significantly enhanced critical current densities in MgB$_2$ tapes made by a scaleable nanocarbon addition route," *Appl. Phys. Lett.*, vol. 88, p. 072502, 2006.

[26] Sang, L. N.; Li, Z.; Yang, G. S.; Yue, Z. J.; Liu, J. X.; Cai, C. B.; Wu, T.; Dou, S. X.; Ma, Y. W.; Wang, X. L., "Pressure effects on iron-based superconductor families: Superconductivity, flux pinning and vortex dynamics," *Materials Today Physics,* vol. 19, p. 100414, 2021.

[27] Flores-Livas, J. A.; Boeri, L.; Sanna, A.; Profeta, G.; Arita, R.; Eremets, M., "A perspective on conventional high-temperature superconductors at high pressure: Methods and materials," *Physics Reports,* vol. 856, p. 1–78, 2020.

[28] Badding, J. V., "High Pressure synthesis, charaterization, and tunning of solid state materials," *Annu. Rev. Mater. Sci.,* vol. 28, p. 631–58, 1998.

[29] Karpinski, J.; Zhigadlo, N.D.; Katrych, S.; Bukowski, Z.; Moll, P.; Weyeneth, S.; Keller, H.; Puzniak, R.; Tortello, M.; Daghero, D.; Gonnelli, R.; Maggio-Aprile, I.; Fasano, Y.; Fischer, O.; Rogacki, K.; Batlogg, B., "Single crystals of LnFeAsO$_{1-x}$F$_x$ (Ln = La, Pr, Nd, Sm, Gd) and Ba$_{1-x}$Rb$_x$Fe$_2$As$_2$: Growth, structure and superconducting properties," *Physica C: Superconductivity,* vol. 469, pp. 370-380, 2009.

[30] Deng, L.; Bontke, T.; Dahal, R.; Xie, Y.; Gao, B.; Li, X.; Yin, K.; Gooch, M.; Rolston, D.; Chen, T.; Wu, Z.; Ma, Y.; Dai, P.; Chu, C.-W., "Pressure-induced high-temperature superconductivity retained without pressure in FeSe single crystals," *Proc. Natl. Acad. Sci. U.S.A.*, vol. 118, p. e2108938118, 2021.

[31] Morawski, A.; Lada, T.; Paszewin, A.; Przybylski, K., "High gas pressure for HTS single crystals and thin layer technology," *Supercond. Sci. Technol.,* vol. 11, p. 193, 1998.

[32] Paranthaman, M.; Chakoumakos, B. C., "Crystal Chemistry of HgBa$_2$Ca$_{n-1}$Cu$_n$O$_{2n+2+\delta}$ (n = 1, 2, 3, 4) Superconductors," *J. Solid State Chem.,* vol. 122, pp. 221-230, 1996.

[33] Podlesnyak, A.; Mirmeistein, A.; Bobrovskii, V.; Voronin, V.; Karkin, A.; Zhdakhin, I.; Goshchitskii, B.; Midberg, E.; Zubkov, V.; Dyachkova, T.; Khlybov, E.; Genoud, J.-Y.; Rosenkranz, S.; Fauth, F.; Henggeler, W.; Furrer, A., "New elaboration technique, structure and physical properties of infinite-layer Sr$_{1-x}$Ln$_x$CuO$_2$ (Ln = Nd, Pr)," *Physica C: Superconductivity,* vol. 258, pp. 159-168, 1996.

[34] Karpinski, J.; Schwer, H.; Mangelschots, I.; Conder, K.; Morawski, A.; Lada, T.; Paszewin, A., "Single crystals of Hg$_{1-x}$Pb$_x$Ba$_2$Ca$_{n-1}$Cu$_n$O$_{2n+2+\delta}$ and infinite-layer CaCuO$_2$. synthesis at gas pressure 10 kbar, properties and structure," *Physica C: Superconductivity,* vol. 234, pp. 10-18, 1994.

[35] Tkachenko, O.; Morawski, A.; Zaleski, A. J.; Przyslupski, P.; Dietl, T.; Diduszko, R.; Presz, Werner-Malento, A. K., "Synthesis, Crystal Growth and Epitaxial Layer Deposition of FeSe$_{0.88}$ Superconductor and Other Poison Materials by Use of High Gas Pressure Trap System," *J. Supercond. Nov. Magn.,* vol. 22, pp. 599-602, 2009.

[36] Durrell, J. H.; Eom, C.-B.; Gurevich, A.; Hellstrom, E. E.; Tarantini, C.; Yamamoto, A.; Larbalestier, D. C., "The behavior of grain boundaries in the Fe-based superconductors," *Rep. Prog. Phys.,* vol. 74, p. 124511, 2011.

[37] Pallecchi, I.; Eisterer, M.; Malagoli, A.; and Putti, M., "Application potential of Fe-based superconductors," *Supercond. Sci. Technol.,* vol. 28, p. 114005, 2015.

[38] Kametani, F.; Polyanskii, A.A.; Yamamoto, A.; Jiang, J.; Hellstrom, E.E.; Gurevich, A.; Larbalestier, D.C.; Ren, Z.A.; Yang, J.; Dong, X.L., "Combined microstructural and magneto optical study of current flow in Nd and Sm Fe-oxypnictides," *Supercond. Sci. Technol.,* vol. 22, p. 015010, 2009.

[39] Singh, S.J.; Shimoyama, J.-I.; Yamamoto, A.; Ogino, H.; Kishio, K., "Significant enhancement of the intergrain coupling in lightly F-doped SmFeAsO superconductors," *Supercond. Sci. Technol.*, vol. 26, p. 065006, 2013.

[40] Yamamoto, A.; Jiang, J.; Kametani, F.; Polyanskii, A.; Hellstrom, E.; Larbalestier, D.; Martinelli, A.; Palenzona, A.; Tropeano. M.; and Putti, M., "Evidence for electromagnetic granularity in polycrystalline Sm1111 iron-pnictides with enhanced phase purity," *Supercond. Sci. Technol.*, vol. 24, p. 045010, 2011.

[41] Guowei, Li; Zhang, B.; Baluyan, T.; Rao, J.; Wu, J.; Novakova, A. A.; Rudolf, P.; Blake, G. R.; Groot, R. A.; and Palstra, T. T., "Metal−Insulator Transition Induced by Spin Reorientation in $Fe_7Se_8$," *Inorg. Chem.*, vol. 55, p. 12912–12922, 2016.

[42] Iida, K.; Hänisch, J.; Yamamoto, A., "Grain boundary characteristics of Fe-based superconductors," *Supercond. Sci. Technol.*, vol. 33, p. 043001, 2020.

[43] Feldmann, D. M.; Holesinger, T. G.; Feenstra, R.; Larbalestier, D. C., "A Review of the Influence of Grain Boundary Geometry on the Electromagnetic Properties of Polycrystalline $YBa_2Cu_3O_{7-x}$ Films," *J. Am. Ceram. Soc.*, vol. 91, p. 1869, 2008.

[44] Singh, S. J.; Bristow, M.; Meier, W. R.; Taylor, P.; Blundell, S. J.; Canfield, P. C.; Coldea, A. I., "Ultrahigh critical current densities, the vortex phase diagram, and the effect of granularity of the stoichiometric high-$T_c$ superconductor $CaKFe_4As_4$," *Phys. Rev. Materials,* vol. 2, p. 074802, 2018.

[45] Chen, D. X.; Pardo, E.; and Sanchez, A., "Demagnetizing factors of rectangular prisms and ellipsoids," *IEEE Trans. Magnet.*, vol. 38, pp. 1742-1752, 2002.

[46] R. Prozorov, R.; Kogan, V. G., "Effective Demagnetizing Factors of Diamagnetic Samples of Various Shapes," *Phys. Rev. Applied,* vol. 10, p. 014030 , 2018.

[47] Tsurkan, V.; Deisenhofer, J.; Gunther, A.; Kant, Ch.; Nidda, H.-A. K.; Schrettle, F.; and Loidl, A., "Physical properties of $FeSe_{0.5}Te_{0.5}$ single crystals grown under different conditions," *Eur. Phys. J. B*, vol. 79, p. 289–299, 2011.

[48] Bean, P.C., "Magnetization of high-field superconductors," *Rev. Mod. Phys.*, vol. 36, p. 31, 1985.

[49] Anderson, P. W.; Kim, Y. B., "Hard Superconductivity: Theory of the Motion of Abrikosov Flux Lines," *Rev. Mod. Phys.*, vol. 36, p. 39, 1964.

[50] Chen, N.; Liu, Y.; Ma, Z.; Li, H.; Hossain, M. S., "Enhancement of superconductivity in the sintered $FeSe_{0.5}Te_{0.5}$ bulks with proper amount of Sn addition," *J. Alloys Compd.*, vol. 633, p. 233–236, 2015.

[51] Zignani, C. F.; Marzi, G. De; Corato, V.; Mancini, A.; Vannozzi, A.; Rufoloni, A.; Leo, A.; Guarino, A.; Galluzzi, A.; Nigro, A.; Polichetti, M.; Corte, A.; Pace, S.; Grimaldi, G., "Improvements of high-field pinning properties of polycrystalline Fe(Se, Te) material by heat treatments," *J. Mater. Sci.*, vol. 54, p. 5092–5100, 2019.

[52] Galluzzi, A.; Buchkov, K.; Tomov, V.; Nazarova, E.; Leo, A.; Grimaldi, G.; Polichetti, M., "High Pinning Force Values of a Fe(Se, Te) Single Crystal Presenting a Second Magnetization Peak Phenomenon," *Materials,* vol. 14, p. 5214, 2021.

[53] Dew-Hughes, D., "Flux pinning mechanisms in type II superconductors," *Philos. Mag.*, vol. 30, pp. 293-305, 1974.

[54] Eisterer, M., "Magnetic properties and critical currents of $MgB_2$," *Supercond. Sci. Technol.*, vol. 20, p. R47–R73, 2007.

[55] Kováč, P.; Hušek, I.; Melišek, T., "$MgB_2$ Composite Superconductors Made by Ex-Situ and In-Situ Process," *Advances in Science and Technology,* vol. 47, pp. 131-136, 2006.